\documentclass[12pt]{article}
\usepackage{amsmath,amssymb,bm,graphicx}
\usepackage{color} 

\newcommand{\pslash}{\not \! p}

\begin{document}

\begin{flushright}
\end{flushright}

\vskip 0.5 truecm

\begin{center}
{\Large{\bf 
Remark on neutrino oscillations }}
\end{center}
\vskip .5 truecm
\begin{center}
\bf { Kazuo Fujikawa }
\end{center}

\begin{center}
\vspace*{0.4cm} 
{\it {Interdisciplinary Theoretical and Mathematical Sciences Program (iTHEMS),\\
RIKEN, Wako 351-0198, Japan}
}
\end{center}
\makeatletter
\makeatother


\begin{abstract}
The oscillations of  ultra-relativistic neutrinos are  
realized by the propagation of  assumed zero-mass on-shell neutrinos with the speed of  light in vacuum combined with the phase modulation by the small mass term $\exp[-i(m^{2}_{\nu_{k}}/2|\vec{p}|)\tau]$ with a time parameter $\tau$.  This picture is realized  in the first quantization by the mass expansion and in field theory by
the
use of $\delta(x^{0}-y^{0}-\tau) 
\langle 0|T^{\star}\nu_{L k}(x)\overline{\nu_{L k}(y)}|0\rangle$
 with the neutrino mass eigenstates $\nu_{L k}$ and  a finite positive $\tau$ after the contour integral of the propagating neutrino energies. 
 By noting that the conventional detectors are insensitive to  neutrino masses, the measured energy-momenta of the initial and final states with assumed zero-mass neutrinos are conserved.  The propagating neutrinos  preserve  the three-momentum in this sense but  the energies of the massive neutrinos are conserved up to  uncertainty relations and thus leading to oscillations.   Conceptual complications in the case of Majorana neutrinos due to the charge conjugation in $d=4$ are also discussed.
\end{abstract}


\section{Neutrino oscillations and mass expansion}

The phenomenon of neutrino oscillations \cite{ MNS, Pontecorvo-0,Pontecorvo-1} is fundamental to measure the small neutrino masses, and it would be disastrous if the different formulations should lead to different neutrino masses.  
 If one writes the neutrino mixing 
with the PMNS matrices $U^{\alpha k}$ 
\begin{eqnarray}\label{naive relation1}
|\nu_{\alpha}\rangle = \sum_{k}{U^{\alpha k}}^{\star} |\nu_{k}\rangle,
\end{eqnarray}
where  $|\nu_{k}\rangle$ are the mass eigenstates which diagonalize the neutrino mass matrix, and the flavor eigenstates $|\nu_{\alpha}\rangle$, $(\alpha=e, \mu, \tau)$,  are related to each other by the above mixing formula. We define the charged lepton flavor eigenstates by the mass eigenstates. 
 One may start with the production of the flavor eigenstate neutrino $\nu_{\mu}$ in the energetic pion decay 
\begin{eqnarray}\label{pion decay}
\pi^{+} \rightarrow \mu^{+} + \nu_{\mu},
\end{eqnarray}
for example, by measuring $\pi^{+}$ and $\mu^{+}$, and the neutrinos thus produced propagate toward the detector in the oscillation experiment; the direction of each mixed  neutrinos may not necessarily be in the exact  specific direction considering the accuracy of the measurements of $\pi^{+}$ and $\mu^{+}$. The oscillation is observed in each direction of the mixed neutrinos. We analyze the Dirac neutrinos in the main part of the present paper for simplicity, and the case of the Majorana neutrinos, which are constrained by the complications of the Majorana fermions in $d=4$, shall be discussed in Appendix.  
It is known that the relation \eqref{naive relation1}, if interpreted as a superposition of on-shell mass eigenstates with  {\em identical} three-momentum,  leads to the standard oscillation  formula \cite{Pontecorvo}
\begin{eqnarray}\label{Pontecorvo1} 
|\langle\nu_{\beta}(0)|\nu_{\alpha}(t)\rangle|^{2} &=& |\sum_{k}U^{\beta k}\exp[i\vec{p}\cdot \vec{x}-i\sqrt{\vec{p}^{2}+m_{\nu_{k}}^{2}}t](U^{\dagger})^{k\alpha}|^{2}\nonumber\\
&=&|\sum_{k}U^{\beta k}\exp[-i\frac{m_{\nu_{k}}^{2}}{2|\vec{p}|}t](U^{\dagger})^{k\alpha}|^{2},
\end{eqnarray}
where $t=L$, the neutrino propagation distance,  is assumed together with $|\vec{p}|^{2}\gg m^{2}_{\nu_{k}}$.  
 It is known also that the identical energy assumption of neutrinos, instead of the identical three-momentum assumption in the above derivation,   gives  essentially the same formula \cite{Akhmedov}. In this paper we want to understand how the oscillation formula is robust against various ways to derive it.
 
 We first mention the idea of the wave packet of neutrinos  in the first quantization formalism \cite{Kayser}.
If one of the mass eigenstates in the neutrino $\nu_{\mu}$, for example $\nu_{1}$, should be identified immediately after the pion decay \eqref{pion decay} such a mass eigenstate due to the reduction of quantum states  would propagate without oscillations, although the flavor change $\beta\rightarrow\alpha$ would be induced by the (inverse) mixing in 
\eqref{naive relation1}. The actual values of neutrino masses are, however, very small and thus the specification of the mass of one of neutrinos is practically impossible; in fact, as is explained later,  the neutrino masses need to be not measured by the conventional detectors to measure the neutrino oscillations.  To treat the un-identified neutrino mass eigenstates consistently,  Kayser  \cite{Kayser} suggested the idea of the wave packet of  particles involved, such as $\nu_{\mu}$ in \eqref{pion decay}. The wave packet is more generally understood as a means to incorporate the semi-classical aspects of neutrino oscillations into quantum mechanics in a consistent manner, and it has been successfully incorporated in the field theoretical formulations  \cite{Akhmedov, Giunti-Kim1993, Grimus1996, Giunti-Kim1998, Grimus1999, Beuthe} and a related quantum mechanical formulation \cite{Rich}. It has been shown 
also that the kinematics of the neutrino production generally implies the mass dependence of the neutrino momentum such as $\vec{p}=\vec{p}(m^{2}_{\nu_{k}})$ and that the mass-dependence of the momentum depends how they are produced \cite{Winter, Giunti-Kim2001}; for example, the two-body decay $\pi^{+}\rightarrow \mu^{+}+\nu$ or other neutrino production processes.

One may consider the propagating phase of a flavor eigenstate in vacuum determined by the phase of the mass eigenstates initially located at $(t,\vec{x})=(0,0)$
 \begin{eqnarray}
 |\nu_{\alpha}(t, \vec{x})\rangle = \sum_{k}{U^{\alpha k}}^{\star}\exp{[i\phi(t, \vec{x}; m_{{\nu}_{k}})]}|\nu_{k}(0,0)\rangle 
 \end{eqnarray}
 by assuming that the neutrino masses are very small and  thus the measured neutrinos are ultra-relativistic, in accord with experimental facts which imply the mass differences $\Delta m^{2}$ on the order of $(10^{-2} eV)^{2}$. One then obtains the Lorentz invariant 
 \begin{eqnarray}\label{neutrino phase}
 \phi(t, \vec{x}; m_{{\nu}_{k}})
 &=&\vec{p}(m^{2}_{{\nu}_{k}})\vec{x} -E(\vec{p}(m^{2}_{{\nu}_{k}}), m^{2}_{{\nu}_{k}}) t\nonumber\\
 &=&\vec{p}(0)\vec{x} -E(\vec{p}(0),0)t +{m^{2}_{{\nu}_{k}}}{\vec{p}^{\ \prime}(0)}\cdot (\vec{x}-\vec{v}_{g}t)-\frac{1}{2}\frac{{m}_{{\nu_{k}}}^{2}}{|\vec{p}(0)|}t +O({m}_{{\nu_{k}}}^{4})
 \end{eqnarray}
 where 
 \begin{eqnarray}
 \vec{p}(m^{2}_{{\nu}_{k}})&=&\vec{p}(0)+ {m^{2}_{{\nu}_{k}}}{\vec{p}^{ {\ \prime}}(0)} + O({m}_{{\nu_{k}}}^{4}),\nonumber\\
  E(\vec{p}(m^{2}_{{\nu}_{k}}), m^{2}_{{\nu}_{k}})&=&\sqrt{\vec{p}(m^{2}_{{\nu}_{k}})^{2}+ {m^{2}_{{\nu}_{k}}}},  \nonumber\\
&=& E(\vec{p}(0), 0)+ {m^{2}_{{\nu}_{k}}}{\vec{p}^{\ \prime}(0)}\cdot \vec{v}_{g}+\frac{1}{2}\frac{{m}_{{\nu_{k}}}^{2}}{|\vec{p}(0)|} +O({m}_{{\nu_{k}}}^{4}) 
 \end{eqnarray} 
 with 
 \begin{eqnarray}\label{group velocity}
 E(\vec{p}(0), 0)=\sqrt{\vec{p}(0)^{2}}, \ \ \ \ 
 \vec{v}_{g}=\frac{\partial E(\vec{p}(0),0)}{\partial \vec{p}(0)}=\frac{\vec{p}(0)}{|\vec{p}(0)|}.
 \end{eqnarray}
 We assume that the momentum $\vec{p}(0)$ is common to all the mass eigenstates of neutrinos. 
The velocity $\vec{v}_{g}$  is the group velocity of the propagating (now regarded as massless) neutrinos. 
The basic assumption of the wave packet (although we do not write an explicit form of the wave packet following  the analysis by Giunti and Kim \cite{Giunti-Kim2001}) is that the neutrinos are concentrated at the center of the wave packet (see also \cite{Akhmedov})
 \begin{eqnarray}\label{center of a wave packet}
 \vec{x}-\vec{v}_{g}t\simeq0.
 \end{eqnarray}
 The neutrino wave packets are essentially the spreading of the initial and final state weak vertices, since the neutrinos rarely interact with surrounding materials. We assume that the geometrical spreads of the weak vertices, which are large in the microscopic sense  so that the spread of the neutrino momentum is negligibly small by the uncertainty principle, but still the geometrical spreads are very small compared with the macroscopic distance between the two weak vertices.
In other words, all the neutrino mass eigenstates are assumed to be measured at $\vec{x}-\vec{v}_{g}t \simeq 0$ or at a finite distance away from $0$, then  the term ${m^{2}_{{\nu}_{k}}}{\vec{p}^{\ \prime}(0)}\cdot (\vec{x}-\vec{v}_{g}t)$ is much smaller than $-\frac{1}{2}\frac{{m}_{{\nu_{k}}}^{2}}{|\vec{p}(0)|}t$ in \eqref{neutrino phase} for large $t$; this is also ensured by the fact that $|{m^{2}_{{\nu}_{k}}}{\vec{p}^{\ \prime}(0)}|$ is of about equal magnitude to $\frac{1}{2}\frac{{m}_{{\nu_{k}}}^{2}}{|\vec{p}(0)|}$, which is confirmed to be the case. The small quantities with $O(m^{4}_{\nu_{k}})$ or higher powers in the neutrino mass are neglected.

 One thus measures the oscillations caused by next to the last term of \eqref{neutrino phase} with the common momentum factor
 \begin{eqnarray}\label{Oscillation factor}
\sum_{k}U^{\beta k}\exp[-i\frac{m_{\nu_{k}}^{2}}{2|\vec{p}(0)|}t](U^{\dagger})^{k\alpha}.
\end{eqnarray}
 When the exponential factor is written in the form  
 \begin{eqnarray}
 -(\sqrt{\vec{p}(0)^{2}+{m^{2}_{{\nu}_{k}}}}- \sqrt{\vec{p}(0)^{2}})t,
 \end{eqnarray}
  \eqref{Oscillation factor} is universal, i.e., depends on the intrinsic properties of the neutrinos  independently  of how the neutrinos were produced. The common factor $\vec{p}(0)\vec{x} -E(0)t$ for each mass eigenstate in \eqref{neutrino phase} does not contribute to the oscillation. 
   We thus recover the standard oscillation formula \eqref{Pontecorvo1} using $t\simeq L$ which arises from  $\vec{x}-\vec{v}_{g}t \simeq 0$ in \eqref{center of a wave packet} with $|\vec{v}_{g}|=1$. 
   Physically, the semi-classical relation $t\simeq L$ with a small error, which is determined by the conventional detectors, is not influenced by the neutrino masses.  In the above analysis, 
 we chose the vanishing masses of propagating neutrinos as the fiducial values and the observed neutrinos are assumed to be essentially massless. 
   
 Alternatively, if one  assumes the time-to-distance conversion $t=L$ to be a valid ansatz for  massless on-shell neutrinos \footnote{This idea of the time-to-distance conversion was criticized in \cite{Akhmedov}. To avoid the criticism, we use this idea only for the on-shell massless neutrinos in \eqref{neutrino phase} and also in the field theoretical amplitude discussed  in the next section. See also \cite{Volobuev1}.}, one would obtain the formula \eqref{Oscillation factor} directly since the term ${m^{2}_{{\nu}_{k}}}{\vec{p}^{\ \prime}(0)}\cdot (\vec{x}-\vec{v}_{g}t)$ then vanishes. Also this picture is consistent with the exact three-momentum conservation induced by the three dimensional integration at both of the initial and final weak vertices, which are implicitly assumed.
We derived the formula \eqref{neutrino phase} by the mass expansion, which is an expansion in terms of a Lorentz scalar quantity of the Lorentz invariant phase $\vec{p}(m^{2}_{{\nu}_{k}})\vec{x} -E(m^{2}_{{\nu}_{k}}) t$; this may imply the Lorentz invariance of the oscillation formula \eqref{Oscillation factor} \cite{Akhmedov}. 
 The present picture may agree with our intuitive understanding of neutrino oscillations; the ultra-relativistic neutrinos propagate with the speed of  light $\vec{x}-\vec{v}_{g}t=0$ in vacuum for a measured momentum $\vec{p}(0)$,  and the effects of the small mass differences $-\frac{1}{2}\frac{m^{2}_{{\nu}_{k}}}{|\vec{p}(0)|}t$ are measured by oscillations in  vacuum \footnote{The oscillations in the dense medium are not considered here.}. The ratio $-\frac{1}{2}\frac{m^{2}_{{\nu}_{k}}}{|\vec{p}(0)|}$ provides an important quantity in this analysis, namely, it needs to be very small and not measured  by the conventional detectors; this constraint, namely,  not measurable by conventional detectors,   generally arises because of the  energy non-conservation in neutrino oscillations (or by an analysis of energy-time uncertainty relations).  
  In the next section on the Feynman amplitude approach to  neutrino oscillations we discuss how the same criterion arises.

 \section{Feynman amplitude approach}
 
To understand the oscillation phenomena in a field theoretical formulation, one may start with an extension of the Standard Model. The  leptonic sector is given by
 \begin{eqnarray}\label{leptonic Lagrangian}
 {\cal L}_{leptons}(x) &=& \left(\overline{e}, \overline{\mu},\overline{\tau}\right)[i\gamma^{\alpha}\partial_{\alpha} 
-\left(\begin{array}{ccc}
m_{e}&0&0\\
0&m_{\mu}&0\\
0&0&m_{\tau}
\end{array}\right)]
\left(\begin{array}{c}
e\\
\mu\\
\tau
\end{array}\right)
\nonumber\\
&+&
\left(\overline{\nu}_{1}, \overline{\nu}_{2},\overline{\nu
}_{3}\right)[i\gamma^{\alpha}\partial_{\alpha} 
-\left(\begin{array}{ccc}
m_{\nu_{1}}&0&0\\
0&m_{\nu_{2}}&0\\
0&0&m_{\nu_{3}}
\end{array}\right)]
\left(\begin{array}{c}
\nu_{1}\\
\nu_{2}\\
\nu_{3}
\end{array}\right)
\nonumber\\
&-&\frac{g}{\sqrt{2}} \{ (\overline{\nu}_{1},\overline{\nu}_{2},\overline{\nu}_{3})_{L}[U^{\dagger}\gamma^{\alpha} W_{\alpha}]\frac{(1-\gamma_{5})}{2}
\left(\begin{array}{c}
e \\
\mu\\
\tau
\end{array}\right) + h.c.\}
\end{eqnarray}
with a $3\times 3$ unitary mixing matrix U in \eqref{naive relation1}. We ignore the neutral current and electromagnetic  interactions.   
 All the particles belong to respective  mass eigenstates and the lowest order Feynman amplitudes (using  the Fermi approximation) are well-defined without infrared singularities. 
One can confirm that the conventional tree-level Feynman amplitude,  which consists  of the production 
and detection weak vertices connected by the Feynman propagator of massive neutrinos, does not give rise to the neutrino oscillations when one integrates over all the values of the weak  interaction points $x^{\mu}$ and $y^{\mu}$ with $x^{\mu}>y^{\mu}$ as well as $y^{\mu}>x^{\mu}$ in the Fermi approximation of weak interactions we work.  This integration   over interaction points, which incorporates backward  moving off-shell anti-neutrinos as  well as  forward moving off-shell neutrinos,  preserves the energy-momentum precisely at each interaction point.  We thus have no time scale to measure oscillations which are related to the time translation non-invariance.  We shall demonstrate below that the Feynman amplitude with only a part of the forward on-shell neutrino  propagator reproduces the neutrino oscillation amplitude \eqref{Pontecorvo1} by  preserving  the measured overall energy-momentum up to uncertainty relations. Feynman rules are used to specify the quantum mechanically allowed couplings.
 
  The Feynman amplitude approach to neutrino oscillations has been discussed in the field theoretical formulation  by  Kobzarev et al, \cite{Kobzarev}, Grimus and Stockinger 
\cite{Grimus1996}, Giunti, Kim and Lee \cite{Giunti-Kim1998} who emphasized the wave packets, and  using plane waves by  Egorov and Volobuev  \cite{Volobuev1, Volobuev0, Volobuev},  among others. In the latter approach \cite{Volobuev1,Volobuev0, Volobuev}, the (effective)  limit 
  $x^{0}-y^{0}\rightarrow \infty$ was considered in the propagating neutrinos of the form
$ \delta(x^{0}-y^{0}- \tau) \langle T^{\star}\nu_{L}^{l}(x)\overline{\nu_{L}^{k}(y)}\rangle$ using a generalization of the Grimus and Stockinger  theorem 
\cite{Grimus1996} and thus achieving the on-shell condition of all the propagating neutrinos. They emphasized that the momentum space Feynman-like amplitude thus defined produces the oscillation amplitude \cite{Volobuev1,Volobuev0, Volobuev} and  the probability interpretation of the oscillation amplitude  is justified based on the probability interpretation of the  conventional  Feynman amplitude. They note the simplicity of their formulation compared to those of the past formulations such as  \cite{Grimus1996}. We follow the basic ideas of \cite{Grimus1996, Giunti-Kim1998, Volobuev1,Volobuev0, Volobuev}, and we shall simplify the derivation of oscillation amplitudes and add several remarks on the robustness of the amplitude thus derived.

  We write the effective Lagrangian of neutrino processes as  
 \begin{eqnarray}
 {\cal L}_{eff} = \sum_{k}\{ \overline{\nu^{k}(x)} [ i\gamma^{\mu}\partial_{\mu} - M_{\nu_{k}}] \nu^{k}(x) + \overline{J_{R}^{\alpha}(x)}U^{\alpha k}\nu_{L}^{k}(x) + \overline{\nu_{L}^{k}(x)}(U^{\dagger})^{k \alpha}J_{R}^{\alpha}(x)\}
 \end{eqnarray}
 by incorporating the neutrino production and detection processes in the sources $J^{\alpha}_{R}(y)$ and $\overline{J^{\beta}_{R}(x)}$, respectively, which are chosen generally not to be Hermitian conjugate to each other. We  have a generalization of Schwinger's source functions \footnote{Schwinger's source functions in the conventional sense stand for the c-number quantities, whereas we include parts of weak vertices in them. We thus generalize the notations of Schwinger's source functions in the present use and ours are regarded as  short hand notations of the conventional Feynman diagram.} 
\begin{eqnarray}\label{delayed measurement}
 \overline{J^{\beta}_{R}(x)}=\int \frac{d^{4}P_{f}}{(2\pi)^{4}}e^{iP_{f}x}\overline{J^{\beta}_{R}(P_{f})},   \ \ \ J^{\alpha}_{R}(y) = \int \frac{d^{4}P_{i}}{(2\pi)^{4}}e^{-iP_{i}y}J^{\alpha}_{R}(P_{i}). 
\end{eqnarray}
 For example, $\sum_{k}\{\overline{\nu_{L}^{k}(x)}(U^{\dagger})^{k \alpha}J_{R}^{\alpha}(x)\}$ describes the decay $\pi^{+} \rightarrow \mu^{+} +\nu_{\mu}$ and \\ $\sum_{k}\{\overline{J_{R}^{\alpha}(x)}U^{\alpha k}\nu_{L}^{k}(x)\}$ describes the electron  production of $\nu_{e}+n\rightarrow p+ e$. The source functions (in a  generalized  context as above) are expressed in terms of plane waves as in the conventional Feynman amplitudes. 
  In the neutrino oscillation experiments, the macroscopic distance between the production vertex and the detection vertex is one of the main observables.  Following \cite{Volobuev1,Volobuev0, Volobuev}  (see also \cite{Giunti-Kim2001}), we work in the framework of plane waves and consider the  configurations in a 4-dimensional sense  where the neutrino production and  absorption points, which are denoted by $y^{\mu}$ and $x^{\mu}$, respectively, are correlated by a fixed time difference
\begin{eqnarray}
x^{0}=y^{0}+\tau
\end{eqnarray}
with a very large macroscopic $\tau>0$.

 A suitable choice of $\overline{J^{\beta}_{R}(P_{f})}$ and $J^{\alpha}_{R}(P_{i})$  specifies the initial and final systems of the neutrino oscillation experiments as described above.
We then have the oscillation  amplitude for the initial (production) vertex such as 
\begin{eqnarray}
\pi^{+} \rightarrow \mu^{+} +\nu_{\mu}
\end{eqnarray}
and the final (absorption) vertex such as 
\begin{eqnarray}
\nu_{e}+n\rightarrow p+ e
\end{eqnarray}
with the neutrino oscillations communicating $\nu_{\mu} (\alpha=\mu)$ to $\nu_{e}(\beta=e)$. Our proposal is to analyze the conventional Feynman amplitude for the weak process, where the neutrino is exchanged,
\begin{eqnarray}
\pi^{+} + n \rightarrow \mu^{+} + p + e
\end{eqnarray}
 which
is written by the prescriptions of source functions described above with an extra $\delta$-function as
\begin{eqnarray}\label{field theory of neutrino oscillation}
&&\int 
d^{4}x d^{4}y\nonumber\\
&&\hspace{1cm}\times e^{iP_{f}x}\overline{J^{\beta}_{R}(P_{f})} U^{\beta k}\delta(x^{0} - y^{0} -\tau)
\langle T^{\star}\nu_{L k}(x)\overline{\nu_{L l}(y)}\rangle(U^{\dagger})^{l\alpha}J^{\alpha}_{R}(P_{i})e^{-iP_{i}y}\nonumber\\ 
&=&\int d^{4}x d^{4}y\delta(x^{0} - y^{0} -\tau) e^{iP_{f}x}\overline{J^{\beta}_{R}(P_{f})}(\frac{1-\gamma_{5}}{2}) \nonumber\\
&&\hspace{1cm} \times  U^{\beta k}\int \frac{d^{4}p}{(2\pi)^{4}}\left(\frac{i\pslash}{p^{2}-m^{2}_{\nu_{k}}+i\epsilon}\right)\delta_{kl}
 e^{-ip(x-y)}(U^{\dagger})^{l\alpha}J^{\alpha}_{R}(P_{i})e^{-iP_{i}y}\nonumber\\
 &=&\int  dy^{0} (2\pi)^{3}\delta^{3}(P_{f}-P_{i})\overline{J^{\beta}_{R}(P_{f})}(\frac{1-\gamma_{5}}{2})U^{\beta k}
\int \frac{d^{4}p}{(2\pi)^{4}}\left(\frac{i\pslash}{p^{2}-m^{2}_{\nu_{k}}+i\epsilon}\right)\delta_{kl}\nonumber\\
 &&\hspace{1cm}\times (2\pi)^{3}\delta^{3}(p-P_{i})e^{-ip^{0}\tau +iP_{f}^{0}\tau}(U^{\dagger})^{l\alpha}J^{\alpha}_{R}(P_{i})e^{i(P^{0}_{f}-P^{0}_{i})y^{0}}\nonumber\\
&=& (2\pi)^{4}\delta^{4}(P_{f}-P_{i})\overline{J^{\beta}_{R}(P_{f})}(\frac{1-\gamma_{5}}{2}) \nonumber\\
&&\hspace{1cm}\times U^{\beta k}\int \frac{dp_{0}}{2\pi}\left(\frac{i\pslash}{p^{2}-m^{2}_{\nu_{k}}+i\epsilon}\right)\delta_{kl}e^{-ip^{0}\tau +iP_{f}^{0}\tau}(U^{\dagger})^{l\alpha}|_{\vec{p}=\vec{P_{i}}}J^{\alpha}_{R}(P_{i})
\end{eqnarray}
where $P_{i}=q_{\pi}-p_{\mu}$ is the entering four-momentum in the case of the pion decay,  and $P_{f}=p_{p} +p_{e} -p_{n}$ is  the four-momentum of the outgoing final system; $m^{2}_{\nu_{k}}$ stands for the neutrino mass eigenvalue squared. Up to this point, the formula is faithful to what defined by the first line in \eqref{field theory of neutrino oscillation}. 

We now make an approximation. Since the energy resolution of conventional detectors cannot detect neutrino masses, we neglect the possible neutrino mass dependence in $P_{i}$ and $P_{f}$; those four-momenta are written  as if all the propagating  neutrinos are massless.  The summation $\sum_{k}$ over the neutrino masses then operates only on the neutrino  propagators, and the formula \eqref{field theory of neutrino oscillation} is written after the contour integral over the neutrino energy as
\begin{eqnarray}\label{formula of neutrino oscillation}
&&(2\pi)^{4}\delta^{4}(P_{f}-P_{i})\times \nonumber\\
&&\ \ \overline{J^{\beta}_{R}(P_{f})}(\frac{1-\gamma_{5}}{2}) 
\{\sum_{k}
U^{\beta k}\frac{\pslash}{2p^{0}}e^{-ip^{0}\tau +iP_{i}^{0}\tau}(U^{\dagger})^{k\alpha}|_{ p^{0}=\sqrt{\vec{p}^{2}+m^{2}_{\nu_{k}}}, \ \vec{p}=\vec{P_{i}}}\}J^{\alpha}_{R}(P_{i}).
\end{eqnarray} 
 The effect of the energy non-conservation induced by $\delta(x^{0} - y^{0} -\tau)$ is still seen by the presence of the time  parameter $\tau$ in the formula.

We would like to add several comments on the above derivation of the  formula \eqref{formula of neutrino oscillation}. It implies that the Feynman amplitude with the modified Feynman propagator  \cite{Volobuev1,Volobuev0, Volobuev,Fujikawa}  describing only a part of the forward propagating neutrinos
\begin{eqnarray}\label{modified propagator2}
\delta(x^{0} - y^{0} -\tau)
\langle T^{\star}\nu_{L k}(x)\overline{\nu_{L l}(y)}\rangle
\end{eqnarray}
gives rise to the neutrino oscillation probability for large fixed $\tau$, as is seen in \eqref{Pontecorvo2} later; the formula \eqref{formula of neutrino oscillation} gives the conventional  result \eqref{Pontecorvo1} if one assumes the time-to-distance conversion $\tau=L$.
In the present paper, we adopted the propagator of neutrinos \eqref{modified propagator2}
which is the same as in \cite{Volobuev1,Volobuev0, Volobuev} with  a fixed large $\tau$, but  we obtain the {\em  on-shell} condition of neutrinos by performing the contour integral with respect to the neutrino energy \cite{Fujikawa}, instead of taking the (effective) limit $x^{0}-y^{0}\rightarrow \infty$ in \cite{Volobuev1,Volobuev0, Volobuev} with the help of the Grimus and Stockinger  theorem.  By this way we performed the above calculations of the amplitude \eqref{formula of neutrino oscillation} with an approximation stated above, in the lowest order of perturbation. This simplified evaluation was possible since we assumed that the energy-momentum of the initial and  final systems, represented  by $P_{i}$ and $P_{f}$, respectively,  are independent of the neutrino masses because of the limited accuracy of conventional detectors of weak interactions. The neutrino momentum $\vec{p}=\vec{P}_{i}$ also becomes mass independent. By this assumption we were able to take the summation over the massive neutrinos $\sum_{k}$ outside the $\delta$-function as in the final formula \eqref{formula of neutrino oscillation} \footnote{If one takes the summation  $\sum_{k}$ including  $\delta^{4}(P_{f}-P_{i})$, one would be able to describe the sum of three-independent Feynman amplitudes of mass eigenstates  with an extra constraint  $\delta(x^{0} - y^{0} -\tau)$.}.
We are assuming that the neutrinos and other particles contained in the source functions are expressed by plane waves, and the on-shell condition of propagating neutrinos  $p^{0}=\sqrt{\vec{p}^{2}+m^{2}_{\nu_{k}}}$ arises from the contour integral $\int dp^{0}$  containing  the factor $e^{-ip^{0}\tau}$ with $\tau>0$. 

The phase $e^{iP_{i}^{0} \tau}$ in \eqref{formula of neutrino oscillation}
is common  to all the massive neutrinos and thus  neglected in the absolute square of the amplitude. We have
the essential part of the amplitude characteristic to oscillations from \eqref{formula of neutrino oscillation} \cite{Volobuev1, Volobuev0, Volobuev, Fujikawa}
\begin{eqnarray}\label{relativistic oscillation} 
 \sum_{k} U^{\beta k}\frac{i\pslash}{2p^{0}}e^{-ip^{0}\tau}(U^{\dagger})^{k\alpha}|_{p^{0}=\sqrt{\vec{p}^{2}+m^{2}_{\nu_{k}}}, \ \ \vec{p}=\vec{P_{i}}} \ 
\end{eqnarray}
with $\tau=L$. Namely, we recognize the mass differences of the massive neutrinos only by the neutrino oscillations which are caused by the second term on the right-hand side of
\begin{eqnarray} 
e^{-ip^{0}\tau}=e^{-i\sqrt{\vec{p}^{2}}\tau-i (m_{\nu_{k}}^{2}/2|\vec{p}|)\tau -i O(m_{\nu_{k}}^{4})\tau},
\end{eqnarray}
 leading to the flavor-changed oscillating final states recognized by weak interactions as in \eqref{Pontecorvo1}. 

 For the ultra-relativistic neutrinos,  the spin factor
\begin{eqnarray}\label{spin factor}
\frac{\pslash}{2p_{0}}=\frac{1}{2}[\gamma^{0}+\gamma^{l}\frac{p_{l}}{p_{0}}]
\end{eqnarray}
 in \eqref{relativistic oscillation} is regarded to be independent of the neutrino masses since $p_{l}/p_{0}=[p_{l}/|\vec{p}|](1 - (1/2) m^{2}_{\nu_{k}}/|\vec{p}|^{2} + ...)\simeq p_{l}/|\vec{p}|$, and thus $\pslash/(2p_{0})\simeq[\gamma^{0}+\gamma^{l}p_{l}/|\vec{p}|]/2$ is regarded as independent of the neutrino masses.  
 The amplitude \eqref{formula of neutrino oscillation} then contains the well-known oscillating factor in the quantum mechanical formulation \eqref{neutrino phase} \cite{Pontecorvo} with the spin factor $\frac{(1-\gamma_{5})}{2}\frac{\pslash}{2|\vec{p}|}$, which does not influence the $\tau$ or L-dependence of the oscillation probability and thus may be absorbed in the initial and final states. This appearance  of the spin factor is a new aspect of the Feynman diagram approach, although it does not influence oscillations.  We thus have the essential part of the oscillating amplitude 
\begin{eqnarray}\label{Pontecorvo2}
|\sum_{k} U^{\beta k}e^{-ip_{0}\tau}(U^{\dagger})^{k\alpha}|_{p_{0}=\sqrt{\vec{p}^{2}+m^{2}_{\nu_{k}}}, \ \ \vec{p}=\vec{P_{i}}}|^{2}
=|\sum_{k} U^{\beta k}e^{-i\frac{m^{2}_{\nu_{k}}}{2|\vec{p}|}\tau}(U^{\dagger})^{k\alpha}|_{\vec{p}=\vec{P_{i}}}|^{2}
\end{eqnarray}
with $\tau=L$. 

The analysis of an explicit connection between $\tau$ and the distance $L=|\vec{x}-\vec{y}|$, namely, the time-to-distance conversion, and the specification of the appropriate extensions of the initial $\vec{y}$ and the final $\vec{x}$ around the points fixed by $\tau=L=|\vec{x}-\vec{y}|$ do not appear in the above formulation \eqref{formula of neutrino oscillation}, although we used $\tau=L$ at several places already. This absence of the analysis of the time-to-distance conversion is analogous to the case of \eqref{Pontecorvo1} and thus we have to remedy the shortcomings, although the direction from $\vec{y}$ to $\vec{x}$ in the present case is specified  by the given common momentum $\vec{p}=\vec{P}_{i}$.  
The  simplest idea may be to assume that the time-to-distance conversion is a valid ansatz in the analysis of neutrino oscillations for effectively massless on-shell neutrinos, since the conventional detectors cannot recognize the neutrino masses. One would then obtain  the desired result from \eqref{Pontecorvo2}  (or \eqref{neutrino phase}) directly. This abstract picture is consistent also with the integration over $y^{0}$, namely, the arbitrariness of the origin of initial time in obtaining \eqref{formula of neutrino oscillation}, and also with the exact conservation of the three-momentum at the two weak vertices following from the integration over $\vec{x}$ and $\vec{y}$. Further discussions on this matter shall be given later.

For the specific {\em two-flavor} case  and the non-diagonal process $\mu\rightarrow e$, for example, the formula \eqref{Pontecorvo2} gives the well-known oscillation probability
\begin{eqnarray}\label{oscillation3}
|\langle e|\mu\rangle|^{2}=(\sin2\theta)^{2}\frac{1}{2}\{1-\cos[\left(\frac{m_{\nu_{1}}^{2}- m_{\nu_{2}}^{2}}{2|\vec{p}|}\right)\tau]\}.
\end{eqnarray}  
The interval of $L(=\tau)$ to measure the oscillations is then specified by  the standard
\begin{eqnarray}\label{oscillation condition}
 \left|\frac{m_{\nu_{1}}^{2}- m_{\nu_{2}}^{2}}{2|\vec{p}|}\right|L =  2\pi 
\end{eqnarray}
depending on the mass difference of neutrinos, and the momentum $\vec{p}$ carried by the (massless) neutrinos which is determined by the measured $\vec{p}=\vec{P_{f}}$; this value is assumed to be independent of neutrino masses.  
The precise  energy-conservation (i.e., time-independence) in \eqref{oscillation3} is given  by  
\begin{eqnarray}\label{energy-conservation}
m_{\nu_{1}}^{2}- m_{\nu_{2}}^{2}=0
\end{eqnarray}
namely, the vanishing oscillation $1-\cos[\left(\frac{m_{\nu_{1}}^{2}- m_{\nu_{2}}^{2}}{2|\vec{p}|}\right)\tau]=0$.
The persistent probability is 
\begin{eqnarray}
|\langle \mu|\mu\rangle|^{2}= 1- (\sin2\theta)^{2}\frac{1}{2}\{1-\cos[\left(\frac{m_{\nu_{1}}^{2}- m_{\nu_{2}}^{2}}{2|\vec{p}|}\right)\tau]\}
\end{eqnarray}
which satisfies  $|\langle \mu|\mu\rangle|^{2}+|\langle e|\mu\rangle|^{2}=1$ in the present case of two-flavor neutrinos. In general, the absolute normalization of the oscillation probability is not well-specified \cite{Akhmedov}, in particular, in the present case multiplying the Feynman amplitude by $\delta(x^{0} - y^{0} -\tau)$ and thus using only a part of the Feynman amplitude, but the specific oscillation probabilities are well-normalized as above. 

\section{Discussion and conclusion}
The observed oscillation of  ultra-relativistic neutrinos is based on the two basic conditions. The first is that the massless {\em on-shell} neutrinos (with various originally small mass eigenstates) propagate for a given momentum$\vec{p}(=\vec{P}_{i})$ from the  position $\vec{y}$ to another $\vec{x}$ with the speed of light in  vacuum 
\begin{eqnarray}\label{speed of light2}
\vec{v}\tau=(\vec{p}/\sqrt{\vec{p}^{2}})\tau =\vec{x}-\vec{y}
\end{eqnarray}
when measured with the conventional detectors which do not recognize the neutrino masses; the present consideration is thus limited to  length scales of the neutrino propagation  approximately within those of the atmospheric neutrino oscillations.  This may be called the semi-classical (particle) aspect. 
The second is that the oscillations are caused by the small phase modulation with the common  momentum factor
\begin{eqnarray}\label{oscillation factor2}
\sum_{k}U^{\beta k}\exp[-i\frac{m_{\nu_{k}}^{2}}{2|\vec{p}|}\tau](U^{\dagger})^{k\alpha}
\end{eqnarray}
such as in \eqref{Pontecorvo2} and \eqref{neutrino phase}. This phase modulation is the  quantum mechanical (wave) effect and supplemented by the subsidiary condition  $\tau=L=|\vec{x}-\vec{y}|$ following from the semi-classical consideration \eqref{speed of light2}. Any formulation satisfying these two conditions \eqref{speed of light2} and \eqref{oscillation factor2} gives the  oscillation formula.   

We assumed in Section 2  that 
 the oscillation formula is, in principle,  applicable to any experiments without referring to the specific  positions $\vec{y}$ and $\vec{x}$ of weak vertices, as  the abstract formulations of  \eqref{neutrino phase} and  \eqref{Pontecorvo2} suggest. The assumption of the time-to-distance conversion was accepted as a valid ansatz for semi-classical on-shell massless neutrinos with the (common)  given momentum $\vec{p}$. In this understanding, the oscillation formulas are valid for any (4-dimensional) configurations  parallel transported from each other with fixed $\vec{p}$, independently of the specification of the origin of time $y^{0}$. The neutrino oscillations are universal phenomena and applicable to any chosen $\vec{x}$ and $\vec{y}$ in the direction of $\vec{p}$ with $\tau=L=|\vec{x}-\vec{y}|$.  The simple Feynman diagram approach in \cite{Volobuev1,Volobuev0, Volobuev} and the  present derivation of the oscillation formula \eqref{formula of neutrino oscillation} may be counted among the schemes based on these assumptions.
 
 Alternatively, one may follow the elaborate wave packet analyses in the past \cite{Akhmedov, Giunti-Kim1993, Grimus1996, Giunti-Kim1998, Grimus1999, Beuthe, Rich}. These analyses may be regarded as clarifying the mechanisms how to satisfy these conditions \eqref{speed of light2} and \eqref{oscillation factor2} including the quantum coherence of neutrinos and the specification of positions $\vec{y}$ and $\vec{x}$. The weak vertices at $\vec{x}$ and $\vec{y}$ are treated naturally in these wave packet pictures, while the treatment of weak vertices are less transparent in the present field theoretical treatment with Lorentz invariant plane waves.  In the wave packet picture, one may assume that 
the semi-classical relation \eqref{speed of light2} is valid for the points $\vec{y}$ and $\vec{x}$ that are by themselves spreading over the three-dimensional domains so that the three-momentum conservation is ensured within the constraints of  uncertainty relations. The spreads of the points $\vec{y}$ and $\vec{x}$ of two weak vertices are still assumed  to be much smaller than the semi-classical distance $L=|\vec{x}-\vec{y}|$ between them; as for an explicit wave packet realization of these conditions see, for example, \cite{Giunti-Kim1998}.  
  In fact, the wave packet formalism is regarded as an attempt to incorporate semi-classical constraints such as \eqref{speed of light2} in the framework of quantum mechanics consistently. In this sense, the justification of the present  Feynman amplitude approach is also given by  the idea of wave packets.

 The on-shell neutrinos with the same three-momentum and  different masses  mean that the energy conservation is {\em not} satisfied in the intermediate states of the oscillation in \eqref{formula of neutrino oscillation} (also in \eqref{neutrino phase}), as is well known \cite{Akhmedov}. 
 The time-dependent  neutrino oscillation  in the case of two flavors in \eqref{oscillation3}, for example,  may be regarded as measuring the effective energy-nonconservation in the intermediate states (with  the notation with  explicit $\hbar$)
 \begin{eqnarray}\label{oscillation criterion2}
\Delta E \tau = 2\pi \hbar 
\end{eqnarray}
where
\begin{eqnarray}
\Delta E= |\sqrt{\vec{p}^{2}+m_{\nu_{1}}^{2}} -\sqrt{\vec{p}^{2}+m_{\nu_{2}}^{2}} |= |\frac{m_{\nu_{1}}^{2}- m_{\nu_{2}}^{2}}{2|\vec{p}|}|
\end{eqnarray}
 standing for the propagating neutrino energy splitting. 
The energy non-conservation is  manifested as {\em time-dependent oscillations}, i.e., the breaking of time translation invariance. But one may regard that the energy non-conservation is superficial since the energy-time uncertainty relation  $\Delta E \tau\geq \hbar/2$ is satisfied naturally although rather marginally by \eqref{oscillation criterion2} with macroscopic $\tau$. This shows that  $\Delta E$ is actually very small and thus measurable only with oscillations, of which effects vanish on average, but not measurable with the conventional detectors.  If $\Delta E$ should be measured by conventional detectors,  the energy-nonconservation would be confirmed by oscillations; this would be a contradiction \footnote{ This contradiction would  also be understood as  the conventional detectors measuring the energies below the uncertainty limits. The oscillation (or the non-invariance of time translation) would not occur if $\Delta E$ should be detected by conventional detectors.}.

Without the delta-functional constraint $\delta(x^{0} - y^{0} -\tau)$, one would have the conventional Feynman amplitude  with the propagating neutrinos
\begin{eqnarray}\label{conventional amplitude}
(\frac{1-\gamma_{5}}{2})
 \{\sum_{k}
U^{\beta k}\left(\frac{i\pslash}{p^{2}-M^{2}_{\nu_{k}}+i\epsilon}\right)\delta_{kl}(U^{\dagger})^{l\alpha}\}
\end{eqnarray}
where  $p=P_{i}$(now exact), and because of Feynman's $i\epsilon$ both the forward propagating neutrinos and the backward propagating anti-neutrinos contribute. In contrast, in our formula \eqref{formula of neutrino oscillation}
only a part of the neutrinos propagating forward in time appear and those anti-neutrinos propagating  backward in time with negative energy  are neglected \footnote{We are assuming that the neutrino is propagating in the oscillation experiment, for simplicity. The anti-neutrino oscillation is measured in addition to examine the CP violation.}. 
With the presence of $\delta(x^{0} - y^{0} -\tau)$ in the measurement of oscillations, the neutrino propagation may be regarded as  a large scale quantum effect.  One may regard that the microscopic CP symmetry of the oscillation amplitude is an indication of the CP symmetry of the original Feynman amplitude; the microscopic CP violation is described by the phase of  the PMNS matrix in the case of three or more leptonic flavors (except for an extra CP violating  $U(1)$ phase even for two flavors in the case of Majorana neutrinos, which is however not measured by oscillation experiments \cite{Bilenky}).

In conclusion, we first re-formulated  the ultra-relativistic neutrino oscillations  using the neutrino mass expansion, which simplifies the formulation substantially  in the framework of the first quantization. The characteristic property of neutrino oscillations is that the conventional detectors cannot recognize the finite neutrino masses, and thus the absence of energy-momentum conservation in a precise sense. These ideas were then applied to  the Feynman amplitude approach to neutrino oscillations with plane waves by constraining the amplitude to a part of forward propagating massive neutrinos. The time-to-distance conversion, which may be justified for effectively massless on-shell neutrinos, was assumed to simplify the analyses of ultra-relativistic neutrino oscillations. A unified picture of ultra-relativistic neutrino oscillations was thus presented. 

\section*{Acknowledgements}
 I  thank I. P. Volobuev for calling their works to my attention and A. Tureanu for  stimulating comments. 
 The present work is supported in part by JSPS KAKENHI (Grant No. JP18K03633).

\appendix

\section{ Majorana neutrinos and charge conjugation}

In this appendix we  briefly summarize the conceptual complications in the case of Majorana neutrinos using firstly Weinberg's model  \cite{Weinberg2} and later the seesaw models \cite{Fukugita}.
The Weinberg's model is defined by an effective
Hermitian Lagrangian \cite{Weinberg2}
\begin{eqnarray}\label{Weinberg's model}
 {\cal L}_{\nu} &=& \overline{\nu_{L}}(x)i\gamma^{\mu}\partial_{\mu}\nu_{L}(x)
-(1/2) \{\nu_{L}^{T}(x)CM_{\nu}\nu_{L}(x) + h.c.\}\nonumber\\
&=& (1/2) \{\bar{\psi} i\gamma^{\mu}\partial_{\mu}\psi(x) - \bar{\psi}(x)M_{\nu}\psi(x) \}
\end{eqnarray}
where $M_{\nu}$ stands for the $3\times3$ diagonalized neutrino mass matrix and we defined
\begin{eqnarray}\label{Majorana neutrino}
\psi(x) \equiv \nu_{L}(x) +C\overline{\nu_{L}}^{T}(x),
\end{eqnarray}
with $C=i\gamma^{2}\gamma^{0}$.
 The field $\psi(x)$ satisfies the classical Majorana condition 
\begin{eqnarray}\label{classical Majorana}
\psi(x)= C\overline{\psi(x)}^{T}
\end{eqnarray}
identically regardless of the choice of $\nu_{L}$. One may define the Majorana fermion by \eqref{Majorana neutrino} together with  the Dirac equation $[ i\gamma^{\mu}\partial_{\mu} - M_{\nu}]\psi(x)=0$. This is the conventional procedure.

In general one cannot define simultaneously a Majorana fermion with well-defined C and P and a Weyl fermion for which C nor P are defined in $d=4$. The conceptual complications are how to define an isolated free Majorana neutrino with well-defined C and P, while weak interactions are described by a chiral fermion. Starting with a Dirac fermion,  one may obtain under the charge conjugation
\begin{eqnarray}
\nu_{L}(x)\rightarrow  C\overline{\nu_{R}(x)}^{T}
\end{eqnarray}
instead of the {\em pseudo-charge conjugation} symmetry $\nu_{L}(x)\rightarrow  C\overline{\nu_{L}(x)}^{T}$ \cite{Fujikawa2} {\em implicit} in the  
fermion \eqref{Majorana neutrino}. If one adopts the pseudo-C symmetry $\nu_{L}(x)\rightarrow  C\overline{\nu_{L}(x)}^{T}$ together with the representation of a Majorana fermion \eqref{Majorana neutrino}, one would encounter various puzzling aspects. For example,  the first expression of the effective Lagrangian \eqref{Weinberg's model} is not invariant under the P operation; also the mass term $\nu_{L}^{T}(x)CM_{\nu}\nu_{L}(x)=\nu_{L}^{T}(x)CM_{\nu}(\frac{1-\gamma_{5}}{2})\nu_{L}(x)$ vanishes under the pseudo-C symmetry. Apparently an idea of the pseudo-C symmetry needs to be understood better \cite{Fujikawa2}.

A way to deal with Majorana neutrinos in a  consistent manner may be to use a general class of {\em seesaw models} \cite{Fukugita}, which contain the equal number of left-handed and right-handed fermions.  One may  
use a Bogoliubov-type transformation to change the definition of the vacuum of the Weyl fermion to the vacuum of the Majorana fermion \cite{Fujikawa4}. The conventional approach gives rise to 
\begin{eqnarray}\label{Naive Majorana}
\psi_{+}(x)&=& \nu_{R}+ C\overline{\nu_{R}}^{T}\nonumber\\
\psi_{-}(x)&=& \nu_{L}- C\overline{\nu_{L}}^{T}
\end{eqnarray}
in the seesaw model, which are Majorana fermions with masses $M_{+}\neq M_{-}$ if one uses the pseudo-C symmetry. But the conventional parity is not well-defined.

After a suitable Pauli-Gursey transformation, which is equivalent to the Bogoliubov transformation but extended easily to three generations, one can rewrite the Majorana fermions as solutions of the seesaw model in the form \cite{Fujikawa4}
\begin{eqnarray}\label{true Majorana}
\psi_{\pm}(x)= \frac{1}{\sqrt{2}}[N(x) \pm C\overline{N}^{T}(x)]
\end{eqnarray}
using Dirac-type  massive fermions $N(x)$, satisfying $[ i\gamma^{\mu}\partial_{\mu} - M_{\pm}]\psi_{\pm}(x)=0$ with $\psi_{\pm}(x)^{C}=\pm\psi_{\pm}(x)$ and the same masses $M_{+}\neq M_{-}$ as in \eqref{Naive Majorana}. The parity is defined by \cite{Fujikawa4}
\begin{eqnarray}
 N(x) \rightarrow i\gamma^{0}N(t,-\vec{x}).
 \end{eqnarray}
 Those fields $\psi_{\pm}(x)$ in \eqref{true Majorana} are the Majorana fermions with well-defined C and P in the conventional sense.

The formal left-handed components $\psi_{L}(x)=\frac{(1-\gamma_{5})}{2}\psi_{-}(x)$ with a $U^{\alpha k}$ {\em modified} by the Pauli-Gursey transformation \cite{Fujikawa4} may be  used  in a model of neutrino oscillations \eqref{formula of neutrino oscillation}. The modification of $U^{\alpha k}$ is cancelled by 
 a modification of $\frac{(1-\gamma_{5})}{2}\psi_{-}(x)$
 induced by the Pauli-Gursey transformation when used in the oscillation formula \eqref{formula of neutrino oscillation}. If $\frac{(1-\gamma_{5})}{2}\psi_{-}(x)$ is measured, $\frac{(1+\gamma_{5})}{2}\psi_{-}(x)$ is recovered by the parity operation.
\\
\\

\end{document}